\documentclass[%
 reprint,
 amsmath,amssymb,
 aps,
]{revtex4-2}

\usepackage{graphicx}
\usepackage{dcolumn}
\usepackage{bm}
\usepackage{hyperref}

\usepackage{xcolor}
\usepackage{float}

\hypersetup{
    colorlinks=true,
    citecolor=blue,
    linkcolor=black,
    urlcolor=blue,
}

\begin{document}

\preprint{APS/123-QED}

\title{Universal Turbulent States of Miscible Two-Component Bose-Einstein Condensates}

\author{Issei Doki}
\email{sf23402r@st.omu.ac.jp}
\affiliation{Department of Physics, Osaka Metropolitan University, 3-3-138 Sugimoto, Sumiyoshi-Ku, Osaka 558-8585, Japan}
\author{Makoto Tsubota}
\email{tsubota@omu.ac.jp}
\affiliation{Department of Physics, Osaka Metropolitan University, 3-3-138 Sugimoto, Sumiyoshi-Ku, Osaka 558-8585, Japan}
\affiliation{Nambu Yoichiro Institute of Theoretical and Experimental Physics (NITEP), Osaka Metropolitan University, 3-3-138 Sugimoto, Sumiyoshi-Ku, Osaka 558-8585, Japan}

\date{\today}

\begin{abstract}
We investigate turbulence in miscible two-component Bose-Einstein condensates confined in a box potential using the coupled Gross-Pitaevskii equations. Turbulence is driven by an oscillating force, causing the components to oscillate either in-phase (co-oscillating) or out-of-phase (counter-oscillating). A parameter measuring component separation (0 for overlap, 1 for full separation) reveals two turbulent states: coupled (the parameter $\sim0$) and decoupled (the parameter $\sim0.5$). Co-oscillating flows transition between these states at a critical interaction strength, while counter-oscillating flows consistently show the decoupled state. A probabilistic model predicts the decoupled state’s parameter as $\sqrt{4/\pi - 1} = 0.523$, consistent with simulations. 
\end{abstract}

\maketitle

\section{\label{sec:level1} Introduction}
Exploring the universal properties of turbulence remains one of the central challenges in physics \cite{Frisch1995,Davidson2004}. Quantum turbulence in atomic Bose-Einstein condensates (BECs) offers a promising platform to address this challenge \cite{Bagnato2016, Bagnato2009, Gauthier2019, Johnstone2019}. Numerical simulations of three-dimensional BECs using the Gross-Pitaevskii (GP) equation have revealed the Kolmogorov spectrum—a hallmark of classical turbulence—suggesting a universality shared between classical and quantum turbulence \cite{Nore1997, Nore1997_2, Kobayashi2005, Kobayashi2005jps, Kobayashi2007, Sasa2011}. 
A notable distinction in quantum turbulence is the presence of quantized vortices, which provide a well-defined structure for studying turbulent cascades. However, experimental confirmation of the Kolmogorov spectrum remains challenging, as direct measurements of the velocity field in BECs have not yet been realized.

An additional strength of BECs is their high degree of experimental controllability. Recent advances in experimental techniques, such as the use of box potentials, have enabled detailed investigations of statistical laws and cascade fluxes in wave turbulence within BECs \cite{Nir2016, Nir2019}. These developments have significantly advanced the study of dynamical scaling and the equation of state for turbulence, shedding new light on universal turbulence laws \cite{Nir2019, Sano2022, Zoran2023, Nir2024}.

Furthermore, BECs can be extended to multi-component systems \cite{Kasamatsu2005}, opening avenues to explore novel turbulence phenomena. A two-component BEC, as the simplest example, consists of two distinct types of gases that interact through inter-component coupling and are described by two macroscopic wavefunctions. These interactions introduce unique dynamics inherent to multi-component systems. For instance, whether the ground state exhibits uniform mixing or phase separation depends on the nature of the inter-component interactions.

In three-dimensional systems, the turbulence of a miscible two-component BEC has been investigated under conditions where the two components flow with a relative velocity \cite{Takeuchi2010, Ishino2011}. It has been found that vortex generation and growth, driven by counter-superflow instability, cause momentum exchange, resulting in "mutual friction" between the superfluids.
These effects give rise to phenomena absent in single-component systems, including unique turbulence cascades and structures, highlighting the distinct physics of multi-component turbulence.

Quantum turbulence in immiscible two-component BECs has also been studied in three dimensions \cite{Kadokura2024}.

A key question is whether universal states and characteristic quantities specific to such multi-component turbulence exist. Addressing this question is of great importance, as it may reveal fundamental principles that govern turbulence in multi-component systems.

In this study, we aim to uncover universal turbulence states in two-component BECs. By solving the coupled GP equations, we analyze wave turbulence in a box potential where the two components remain miscible in the ground state. Building on a previously studied single-component system \cite{Nir2019}, we extend it to a two-component system and classify the oscillation into two types: co-oscillating, where both components oscillate in phase, and counter-oscillating, where they oscillate out of phase. Numerical simulations were performed for both types to explore their unique turbulent dynamics.
To characterize the degree of separation between the densities of the two components, we introduced a parameter ranging from 0 (complete overlap) to 1 (complete separation). Our findings reveal two distinct turbulent states: a coupled turbulence state with a parameter value of 0, and the other decoupled turbulence state with a parameter value of 0.5. A probabilistic model predicts the decoupled state’s parameter as $\sqrt{4/\pi - 1} = 0.523$, consistent with simulations. 

\section{ Methods}
We study two-component BECs, each described by a macroscopic wave function, $\psi_j(\boldsymbol{r}, t) = \sqrt{n_j(\boldsymbol{r}, t)} e^{i \phi_j(\boldsymbol{r}, t)}$, where $n_j(\boldsymbol{r}, t) = |\psi_j(\boldsymbol{r}, t)|^2$
represents the particle density, \(\phi_j(\boldsymbol{r}, t)\) is the phase, and the index \(j = 1, 2\) denotes the two condensates.
The dynamics of the system are governed by the coupled dimensionless GP equations given by:
\begin{eqnarray}
i\frac{\partial} {\partial t}\psi_j &=&{\bigg(}-\nabla^2+V_j+\sum_{l=1,2} {g_{jl} |\psi_l|^2 }{\bigg)} \psi_j,
\label{eq:GP}\end{eqnarray}
where $g_{jl}$ are the dimensionless coupling constants, and  $V_j$ refers to the potential acting on the $j$-th component.

The length and time scale are normalized by the coherence length $\tilde{\xi}_j = {\hbar}/{\sqrt{2 \tilde{m}_j \tilde{g}_{jj} \tilde{n}_{0j}}}$ and the time  ${\hbar}/{\tilde{g}_{jj} \tilde{n}_{0j}}$ it takes for sound waves to propagate over a distance of \(\tilde{\xi}_j\), where \(\tilde{m}_j\) is the atomic mass and \(\tilde{n}_{0j}\) is the average density of the $j$-th component.
The coupling constant \(\tilde{g}_{jl}\) is expressed as $\tilde{g}_{jl} = {4\pi \hbar^2 \tilde{a}_{jl}}/{\tilde{m}_{jl}}$, where \(\tilde{m}_{jl}\) denotes the reduced mass of the \(j\)-th and \(l\)-th components, and \(\tilde{a}_{jl}\) represents the \(s\)-wave scattering length between the \(j\)-th and \(l\)-th components. In this study, we primarily focus on miscible BECs, which occur when $\sqrt{g_{11} g_{22}} > g_{12}^2$.
However, when $\sqrt{g_{11} g_{22}} < g_{12}^2$, the two components become immiscible, leading to phase separation. To simplify the analysis, we assume symmetric interactions, setting $g_{11} = g_{22} = g \quad \text{and} \quad g_{12} = g_{21} = \gamma g$ where $\gamma < 1$ ensures that the system remains in the miscible regime.

The potential is given as described in Reference \cite{Sano2022,sup1}, 
$V_j = V^{\rm osc}_j + V^{\rm box} + iV^{\rm diss}$.
Here, $V^{\rm osc}_j(z, t)$ represents the oscillating potential, expressed as 
$V^{\rm osc}_j(z, t) = U^\text{s}_j \sin(\omega_{\rm res} t) {z}/{L}$.
$V^{\rm box}(\bm{r})$ denotes a cylindrical box potential with a depth $U^{\text D}$, height $L$, and radius $R$.
The dissipative term $V^{\rm diss}$ is introduced to remove particles outside the cylindrical region.
The parameter $\omega_{\rm res}$ is the resonant frequency of the Bogoliubov wave with a wavelength of $2L$.
 We investigate both co-oscillating flow ($U^\text{s}_1 = U^\text{s}_2$) and counter-oscillating flow  ($U^\text{s}_1 = -U^\text{s}_2$).
 
 We solve Eq.~(\ref{eq:GP}) under periodic boundary conditions, using the pseudo-spectral method with a grid of size $V_{\rm num}=(L_{\rm num})^3=40^3$ and $N_{\rm grid}=128^3$ points for the spatial part, and a fourth-order Runge-Kutta method with a time step of $10^{-3}$ for the temporal part.
We use typical experimental parameters similar to those used in previous experiments on one-component BEC ~\cite{Nir2016, Nir2019}: $g=0.11$, $L=22$, $R=13$, $U^{\rm D}=64$, $\omega_{\rm res} = 0.24$, and vary $\gamma$ to investigate its dependence.
The wavefunctions are normalized such that $\int |\psi_j(\boldsymbol{r}, 0)|^2 \text{d}^3\bm{r} = N_j(0) = 1.1 \times 10^5$, where $N_j(t)$ is the total particle number of the $j$-th component. 
The initial state, which corresponds to the ground state without the oscillating potential (i.e., $V^{\rm for}_j = 0$), was prepared using imaginary-time propagation. To avoid the system being trapped in local solutions during the propagation, independent noise was introduced into each of the two BECs at the beginning of the imaginary-time propagation. In the fully developed turbulence state, the particle number decreases by approximately $1\%$ due to the dissipation term around $t=1000$.

\begin{figure}
\centering
\includegraphics[width=8.6cm]{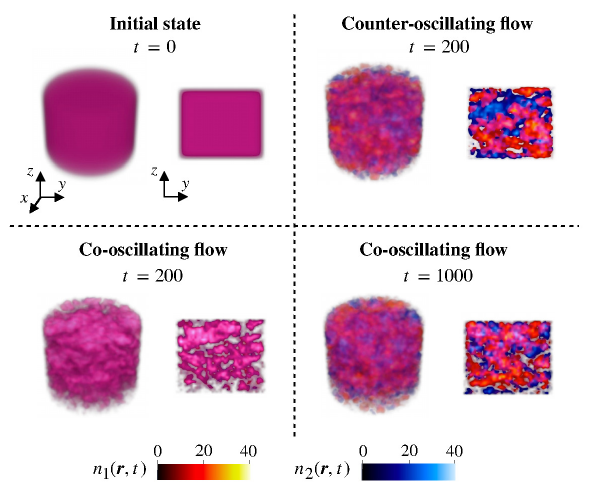}
\caption{(Color online) Dynamics of the density distribution \( n_j(\boldsymbol{r}, t) = |\psi_j(\boldsymbol{r}, t)|^2 \) in three dimensions and on the \( y\text{-}z \) plane at \( x = L_{\rm num}/2 \) for co-oscillating and counter-oscillating flows at \(\gamma = 0.4\). Red and blue represent the densities of components 1 and 2, respectively, while purple indicates regions of overlap. In the co-oscillating flow, the density overlap persists longer due to inter-component interactions. In contrast, in the counter-oscillating flow, decoupling occurs earlier as a result of the opposing oscillations.}
\label{fig:dynamics}
\end{figure}

\begin{figure}
\centering
\includegraphics[width=8.6cm]{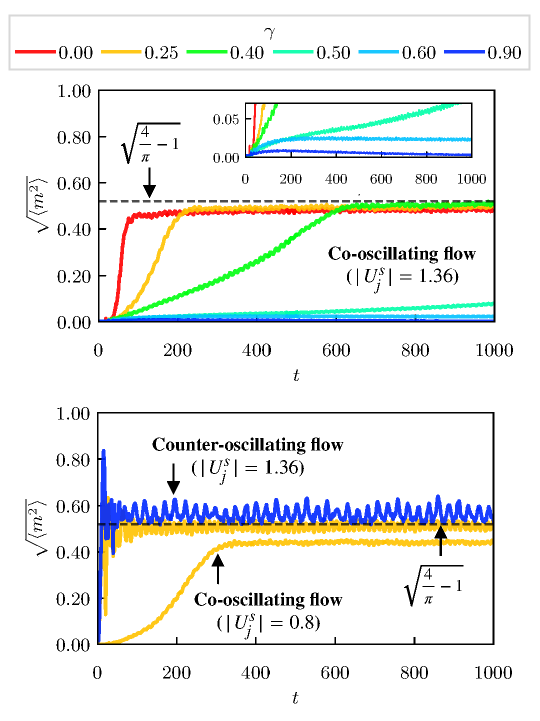}
\caption{(Color online) Time evolution of the density separation parameter \( \sqrt{\langle m^2 \rangle} \) for different \( \gamma \) values and driving amplitudes. The upper panel illustrates the co-oscillating flow with \( |U^{\rm s}_j| = 1.36 \), while the lower panel shows the co-oscillating flow with \( |U^{\rm s}_j| = 0.8 \) and the counter-oscillating flow with \( |U^{\rm s}_j| = 1.36 \). The horizontal black dashed line represents \( \sqrt{4/\pi - 1} \) as a reference. The inset provides a magnified view of \( \sqrt{\langle m^2 \rangle} \), focusing on the vertical axis near zero.}
\label{fig:sep}
\end{figure}

\section{Results}
Figure~\ref{fig:dynamics} show the dynamics of the density distribution $n_j(\boldsymbol r, t) = |\psi_j(\boldsymbol r, t)|^2$ for co-oscillating flow and counter-oscillating flow at $\gamma = 0.4$. Red represents the density of component 1, blue represents the density of component 2, and purple indicates regions where the densities of the two components overlap.
Initially, the two density distributions nearly overlap entirely, with small fluctuations at each position $\bm{r}$. The fluctuations are quantified as $\left| \Delta n \right| = \left| {(n_1(\bm{r},0) - n_2(\bm{r},0))}/{n_0(0)} \right| \leq 0.5\%$, where $n_0(t) = \sum_j {N_j(t)}/(2\pi {R^2} L)$.

Turbulence amplifies fluctuations, leading to the decoupling of the two density distributions. For example, in the co-oscillating flow at $\gamma = 0$, the density distributions begin to separate around $t = 50$.
However, in the co-oscillating flow at $\gamma = 0.4$, the two density distributions remain nearly overlapped even in the turbulent state at $t = 200$, eventually separating as time progresses. This indicates that the timescale for decoupling is significantly longer compared to the $\gamma = 0$ case.
In the co-oscillating flow, whether the two densities overlap or separate is determined by the competition between turbulent fluctuations, which promote decoupling, and inter-component interactions, which work to maintain overlap. This overlap may be driven by mechanisms such as "mutual friction", similar to that observed in counter-superflow instability \cite{Takeuchi2010, Ishino2011, sup2}. 

To quantify this behavior, we adopted the density separation parameter\cite{Johannes2014,Xunda2019}
\begin{eqnarray}  
\sqrt{\langle m^2 \rangle}=\sqrt{\frac{1}{\pi {R^2} L}\int_{\rm box}  m(\boldsymbol r,t)^2{\rm d}^3 \boldsymbol r},  
\label{eq:sep}  
\end{eqnarray}  
where \(m(\boldsymbol r,t)= {(n_1(\boldsymbol r,t)-n_2(\boldsymbol r,t)})/({n_1(\boldsymbol r,t)+n_2(\boldsymbol r,t)})\).  
The integral $\int_{\rm box}{\rm d}^3 \boldsymbol r$ indicates a volume integral performed within the box potential.
 Here, the density separation parameter $\sqrt{\langle m^2 \rangle}$ ranges from $0$ to $1$: $\sqrt{\langle m^2 \rangle}=0$ indicates complete overlap of the two density distributions, while $\sqrt{\langle m^2 \rangle}= 1$ represents no overlap, meaning the two densities are fully separated.

Figure~\ref{fig:sep} show the time evolution of the density separation parameter for different $\gamma$. The upper panel illustrates the co-oscillating flow with \( |U^{\rm s}_j| = 1.36 \), while the lower panel shows the co-oscillating flow with \( |U^{\rm s}_j| = 0.8 \) and the counter-oscillating flow with \( |U^{\rm s}_j| = 1.36 \).

The time evolution of $\sqrt{\langle m^2 \rangle}$ clearly demonstrates that there are only two steady turbulent states, which correspond to two distinct classes.
In the initial state, $\sqrt{\langle m^2 \rangle}$ is almost 0 because BECs are in a miscible ground state.
In the case of co-oscillating flow with $|U^{\rm s}_j| = 1.36$, $\sqrt{\langle m^2 \rangle}$ begins to increase from 0 as turbulence develops. For $\gamma = 0$, corresponding to the single-component case, $\sqrt{\langle m^2 \rangle}$ gradually approaches a value of approximately 0.5 around $t = 100$, after which it reaches a statistically steady turbulent state. This statistically steady turbulent state, characterized by $\sqrt{\langle m^2 \rangle} \approx 0.5$, corresponds to a decoupled turbulent state.
As $\gamma$ increases, $\sqrt{\langle m^2 \rangle}$ still converges to the decoupled turbulent state characterized by $\sqrt{\langle m^2 \rangle} \approx 0.5$, but the relaxation time to reach this state becomes longer.
When $\gamma > 0.6$, $\sqrt{\langle m^2 \rangle}$ initially increases slightly from 0, reflecting the development of turbulent fluctuations.
However, at later times, $\sqrt{\langle m^2 \rangle}$ decreases and approaches 0, after which it reaches a statistically steady turbulent state. This statistically steady turbulent state, characterized by $\sqrt{\langle m^2 \rangle} = 0$, corresponds to a coupled turbulent state.

For a lower driving amplitude of $|U^{\rm s}_j| = 0.8$, $\sqrt{\langle m^2 \rangle}$ also converges to a decoupled turbulent state with a value slightly smaller than 0.5, but the relaxation time is longer compared to the case with $|U^{\rm s}_j| = 1.36$.
In the counter-oscillating flow with $|U^{\rm s}_j| = 1.36$, $\sqrt{\langle m^2 \rangle}$ consistently converges to the decoupled turbulent state with $\sqrt{\langle m^2 \rangle} \approx 0.5$, regardless of the interaction strength $\gamma$. This behavior can be attributed to the opposite directions of the oscillating potential.

The evolution of $\sqrt{\langle m^2 \rangle}$ highlights the presence of two distinct turbulent states: a coupled turbulent state, where $\sqrt{\langle m^2 \rangle}$ converges to 0, and a decoupled turbulent state, where it approaches approximately 0.5. In co-oscillating flows with $|U^{\rm s}_j| = 1.36$, the transition between these states is determined by a critical value of $\gamma$ between 0.5 and 0.6. In contrast, counter-oscillating flows consistently lead to the decoupled turbulent state with $\sqrt{\langle m^2 \rangle} \approx 0.5$, regardless of $\gamma$. Both the coupled and decoupled turbulent states exhibit the Kolmogorov-Zakharov spectrum \cite{Nir2016,Nir2019}, characterized by a power-law dependence $\propto k^{-3.5}$, similar to that observed in single-component systems\cite{sup3}.

To understand why $\sqrt{\langle m^2 \rangle}$  converges to the universal value of 0.5 in the decoupled turbulent state, we employ a simple probabilistic model. We assume that, in this decoupled state, the two BECs behave independently due to turbulent fluctuations. Based on this assumption, the density with temporal fluctuations at a given spatial point is treated as independent random variables.
Specifically, we assume that the density of component 1, $n_1(\boldsymbol r,t)$, is a random variable $X$ and the density of component 2, $n_2(\boldsymbol r,t)$, is a random variable $Y$, both following Gaussian distributions, $f_X$ and $f_Y$, respectively, with a mean value of 0. 
Numerical calculations confirm that the density histograms exhibit the Gaussian distributions\cite{sup4}.

Furthermore, because the BECs are homogeneous in the box potential, $\sqrt{\langle m^2 \rangle}$ can be expressed as $\sqrt{m(\boldsymbol{r},t)^2}$ at a single spatial point, which is statistically steady over time.
The Gaussian distributions are parameterized such that the expected values of the densities satisfy the physical requirement of matching the average density $n_0(t)$.
Therefore, the probability density function can be written as 
\begin{align}
f_X(x)=\frac{2}{\pi n_0}{\rm exp}{\Bigg[}{-{\frac {1}{\pi}{\bigg(}\frac{x}{n_0}{\bigg)}^2}}{\Bigg]},
\label{eq:gauss}
\end{align}
\begin{align}
f_Y(y)=\frac{2}{\pi n_0}{\rm exp}{\Bigg[}{-{\frac {1}{\pi}{\bigg(}\frac{y}{n_0}{\bigg)}^2}}{\Bigg]}.
\label{eq:gauss_y}
\end{align}
Taking the square root of the expected value of ${m(\boldsymbol{r},t)^2}$, we obtain:  
\begin{align}
\sqrt{m(\boldsymbol{r},t)^2}&=\sqrt{\int^\infty_0\int^\infty_0\left(\frac{x - y}{x + y}\right)^2 f_X(x)f_Y(y){\rm d}x{\rm d}y }\\ &= \sqrt{\frac{4}{\pi}-1}.
\label{eq:expected}
\end{align}
This theoretical value $\sqrt{{4}/{\pi}-1}= 0.523$ is a universal one, independent of $n_0$. This result is consistent with numerical simulations, providing a quantitative explanation for the decoupled turbulent state. 

\section{Conclusion}
In this study, we explored the emergence of universal turbulence states in miscible two-component BECs under co-oscillating and counter-oscillating flows by analyzing the density separation parameter $\sqrt{\langle m^2 \rangle}$.
Our numerical simulations revealed the presence of two distinct turbulent states: a coupled turbulent state, where $\sqrt{\langle m^2 \rangle}$ converges to 0, and a decoupled turbulent state, where it approaches approximately 0.5, regardless of the interaction strength, the amplitude of external forces, or the type of forcing. 
In the decoupled turbulent state, a probabilistic model treating the two densities as independent random variables analytically predicts a value of $\sqrt{{4}/{\pi}-1} = 0.523$ for $\sqrt{\langle m^2 \rangle}$, which aligns closely with our numerical results.

In co-oscillating flows, the transition between these states is governed by a critical interaction strength $\gamma_c$, which should depend on the oscillation amplitude $|U^s_j|$. When $\gamma > \gamma_c$, the system enters the coupled state, while for $\gamma < \gamma_c$, it transitions to the decoupled state. It is expected that $\gamma_c$ increases with $|U^s_j|$, as stronger turbulence enhances decoupling. Therefore, constructing a phase diagram of coupled and decoupled turbulent states as a function of $|U^s_j|$ and $\gamma$ could elucidate the interplay between coupling due to "mutual friction" and decoupling due to turbulence, which is a prospect for future studies.
In contrast, counter-oscillating flows consistently result in the decoupled turbulent state, regardless of the value of $\gamma$, as the opposing oscillations enhance the decoupling of the two components.

In summary, we have identified universal turbulence states characteristic of two-component turbulence and introduced a key parameter to characterize these states, marking an important step in understanding two-component quantum turbulence.

\vspace{0.5cm}

\begin{acknowledgments}
{\bf Acknowledgments}\verb|   | The authors are grateful to Nir Navon and Weican Yang for helpful discussions. M.T. acknowledges the support from JSPS KAKENHI Grants No. JP23K03305 and No. JP22H05139. 
\end{acknowledgments}






\setcounter{figure}{0} 
\setcounter{equation}{0} 

\renewcommand\theequation{S\arabic{equation}} 
\renewcommand\thefigure{S\arabic{figure}} 

\onecolumngrid

\section*{Supplemental Material}
\label{SI}

\subsection{Numerical Details of the Potential Term}
We use the potential described in Reference \cite{Sano2022}:
\begin{eqnarray}\label{eq:pot}
V_j = V^{\rm osc}_j+V^{\rm box}+iV^{\rm diss}.
\end{eqnarray}
Here the external forcing potential is 
\begin{equation}
V^{\rm osc }_j (z, t)= U^\text{s}_j\sin(\it \omega_{\rm res} t) \frac{z}{L}\, ,
\label{eq:osc}
\end{equation}
the box potential is 
\begin{align}
V^{\rm box}(\boldsymbol r)=
 \begin{cases}
  \displaystyle   0  & \left( \displaystyle \it |z| \le \frac{L}{\rm 2},\it \sqrt{x^{\rm 2}+\it y^{\rm 2}}\le R\right) \\[4 pt]
  \displaystyle  U_{\rm D} &\,\rm\left(otherwise\right) ,
\label{eq:box}\
\end{cases}\end{align}
and the dissipation term is
\begin{align}
V^{\rm diss}(\boldsymbol r)=
  \begin{cases}
   \displaystyle 0  & \left( \displaystyle \it |z| \le \frac{L+\rm 2 \it \delta}{\rm 2},\it \sqrt{x^{\rm 2}+\it y^{\rm 2}}\le R+\delta \right) \\[4 pt]
   \displaystyle V_{\rm E} & \,\rm\left(otherwise\right) ,
  \end{cases}
\label{eq:diss}
\end{align}
where $R$ is the radius, and $L$ is the length of the cylindrical box potential. 
The parameter $\delta$ is introduced to prevent the dissipation of the small, evanescent-like component of the wavefunction that exists just outside, but close to the edge of the finite-depth box.
We use typical experimental parameters similar to those used in previous experiments on one-component BEC ~\cite{Nir2016, Nir2019}: $L=22$, $R=13$, $U_{\rm D}=64$, $V_{\rm E}=-5.0$, $\delta = 3.0$ and $\omega_{\rm res} = 0.24$.

\subsection{Supporting Evidence for the Role of Mutual Friction in Density Overlap}

In two-component Bose-Einstein condensates (BECs), stronger inter-component repulsive interactions tend to promote phase separation. Even in turbulent states, this tendency is expected to persist. However, the presence of quantized vortices introduces "mutual friction," as demonstrated in previous studies~\cite{Takeuchi2010,Ishino2011}. 
These studies demonstrated that direct current (DC) counter-superflow induces "mutual friction" between the two superfluids, facilitating momentum exchange between them. This momentum exchange becomes more significant with increasing repulsive inter-component interaction strength $\gamma$, resulting in a reduction of the relative velocity of the counterflow.
A similar mechanism is expected to operate also in the alternating current (AC) flows studied here, where "mutual friction" can counteract phase separation and promote coupled turbulence. The numerical results presented in the main text are consistent with this interpretation.

\begin{figure}
\centering
\includegraphics[width=16cm]{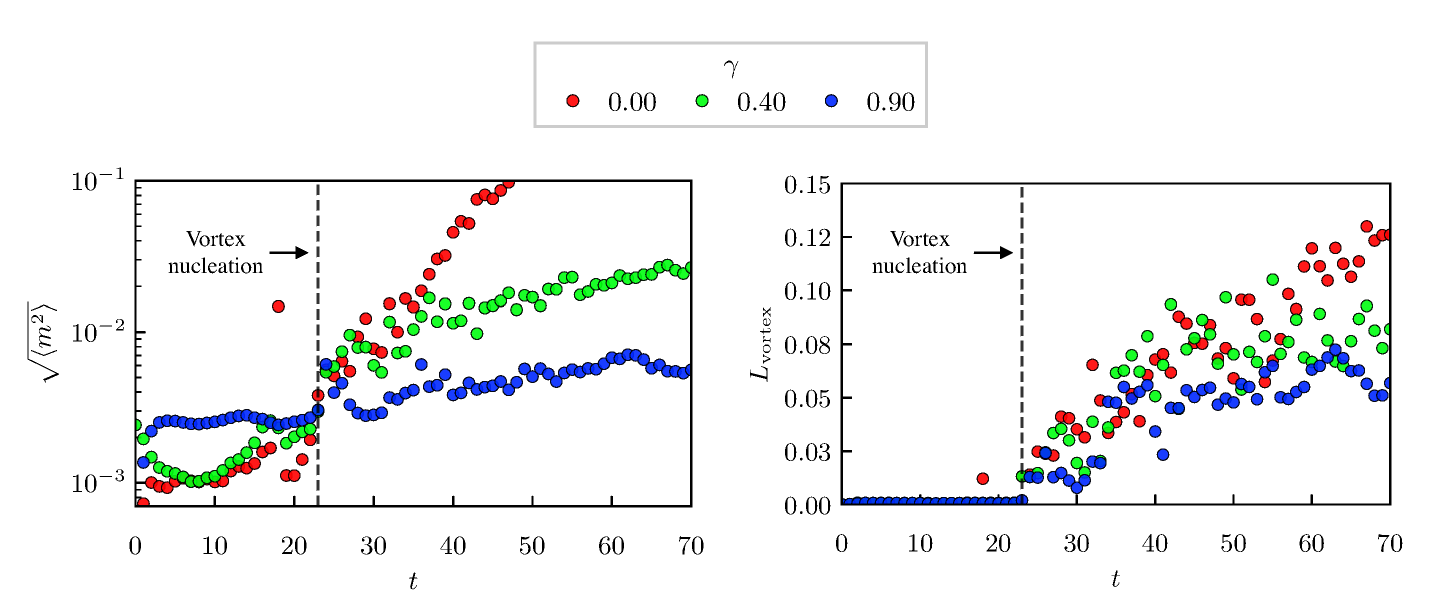}
\caption{The left panel shows the time evolution of \( \sqrt{\langle m^2 \rangle}\) for inter-component interaction strengths \(\gamma = 0.0, 0.4, 0.9\) in a co-oscillating flow with \( U^s_j = 1.36 \), and the right panel shows the evolution of \( L_{\rm vortex} \) for each of these interaction strengths. Both panels focus on the time evolution during the early stages, with vortex formation occurring at \( t \simeq 23 \).
}
\label{f1}
\end{figure}

To further support this result, we analyzed the time evolution of $\sqrt{\langle m^2 \rangle}$ and the vortex line density $L_{\rm vortex}$ during the early stages, as shown in Figure~\ref{f1}.

In the co-oscillating flow with $U^s_j = 1.36$, vortex formation occurs at approximately $t \simeq 23$, regardless of the interaction strength ($\gamma = 0.0, 0.4, 0.9$). Before vortex formation at $t \simeq 23$, an increase in $\gamma$ leads to a rise in $\sqrt{\langle m^2 \rangle}$. In contrast, after vortex formation, "mutual friction" becomes dominant, suppressing further increases in $\sqrt{\langle m^2 \rangle}$ and thereby enhancing the density overlap.

\subsection{Confirmation of Gaussian Distribution in Density Histograms}
To understand why $\sqrt{\langle m^2 \rangle}$  converges to the universal value of 0.5 in the decoupled turbulent state, we employ a simple probabilistic model in the main text.
In the main text, we assume that the density of component 1, with temporal fluctuations at a given spatial point, is represented as a random variable \( X \), while that of component 2 is represented as a random variable \( Y \). Both \( X \) and \( Y \) are assumed to follow Gaussian distributions, \( f_X \) and \( f_Y \), respectively, as described by the following equations:
\begin{eqnarray}
f_X(x)=\frac{2}{\pi n_0}{\rm exp}{\Bigg[}{-{\frac {1}{\pi}{\bigg(}\frac{x}{n_0}{\bigg)}^2}}{\Bigg]}, \ \ f_Y(y)=\frac{2}{\pi n_0}{\rm exp}{\Bigg[}{-{\frac {1}{\pi}{\bigg(}\frac{y}{n_0}{\bigg)}^2}}{\Bigg]}.
\label{eq:gauss}
\end{eqnarray}

\begin{figure}
\centering
\includegraphics[width=8.6cm]{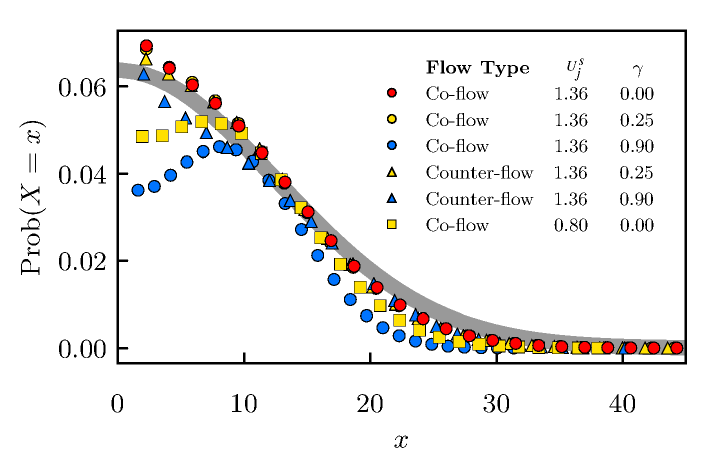}
\caption{The density histograms of component 1 for each type of flow, \( U^s_j \), and inter-component interaction \(\gamma\). Here different marker colors and shapes distinguish the conditions. The gray line represents the Gaussian distribution given by Eq.~(\ref{eq:gauss}) .}
\label{f2}
\end{figure}

In this section, we demonstrate that the density histograms obtained from numerical simulations follow Gaussian distributions. Here, assuming ergodicity based on the homogeneity of the box potential and the stationarity of turbulence, sampling was performed from each numerical grid point within the box during the time interval from 400 to 500 to construct the histograms, as shown in Figure.~\ref{f2}.

As shown in Figure.~\ref{f2}, the histograms for decoupled turbulence exhibit a clear Gaussian shape. However, in the co-oscillating flow case with weak external forcing ($\gamma = 0.25$, $U^s_j = 0.8$), the low-density region deviates from the Gaussian distribution. This indicates that the external forcing is insufficient to fully achieve decoupled turbulence in this scenario. Furthermore, in the coupled turbulence regime ($\gamma = 0.9$, $U^s_j = 1.36$), significant deviations from the Gaussian distribution are observed in the low-density region, highlighting the influence of inter-component interactions on the density fluctuations.

\section{Momentum Spectrum and Particle Flux}
\begin{figure}
\centering
\includegraphics[width=8.6cm]{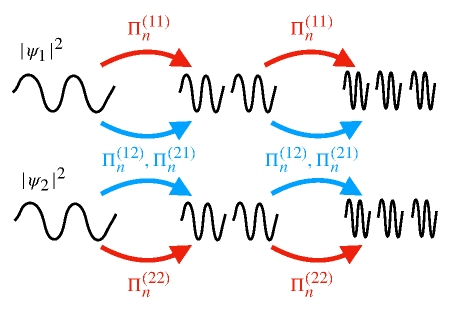}
\caption{Cartoon of the wave turbulent cascade in two-component BECs. Long-wavelength waves break down into shorter-wavelength waves without dissipation, and the shortest-wavelength waves are dissipated as they escape from the box potential. The particle fluxes for each component consist of two terms: the first term, $\Pi^{(jj)}_n$, originates from its nonlinear self-interaction, while the second term, $\Pi^{(jl)}_n \ (j \neq l)$, arises from its nonlinear inter-component interaction. Here, $\Pi^{(jl)}_n \ (j \neq l)$ represents the flux transferring momentum within the same component, not transferring it to the other component because particle types do not transfer between components.}
\label{f3}
\end{figure}

We analyzed momentum distributions and particle fluxes.
The momentum distribution of the $j$-th component defined as \(n_j(\boldsymbol k, t) = |\psi_j (\boldsymbol k, t)|^2\), where \(\psi_j (\boldsymbol k, t)\) is the Fourier component of \(\psi_j (\boldsymbol r, t)\).
The continuity equation of the momentum distribution\cite{Nir2024} is
\begin{eqnarray}
\frac{\partial n_j(\boldsymbol k,t)}{\partial t} 
 =- \sum_{l=1,2}\nabla \cdot \boldsymbol {\Pi}^{(jl)}_n(\boldsymbol{k},t)
+ \mathcal{F}^{(j)}(\boldsymbol{k},t)
+ \mathcal{D}^{(j)}(\boldsymbol{k},t).
\label{eq:bud}\end{eqnarray}
Here, the first term represents the momentum transport term:
\begin{eqnarray}
\nabla \cdot \boldsymbol \Pi^{(jl)}_n(\boldsymbol{k}, t)=2i{\rm Im}{\big[}\mathbb{F}[g_{jl} |\psi_l(\boldsymbol r,t)|^2\psi_j(\boldsymbol r,t)] \psi_j^*(\boldsymbol r,t){\big]},
\label{eq:bud_nl}\end{eqnarray}
the second term corresponds to the forcing term:
\begin{eqnarray}
\mathcal{F}^{(j)}(\boldsymbol{k},t)=2i{\rm Im}{\big[}\mathbb{F}[V^{osc}_j(\boldsymbol r,t) \psi_j(\boldsymbol r,t) ]\psi_j^*(\boldsymbol r,t){\big]},
\label{eq:bud_f}\end{eqnarray}
and the third term denotes the dissipation term:
\begin{eqnarray}
\mathcal{D}^{(j)}(\boldsymbol{k},t)=2i{\rm Im}{\big[}\mathbb{F}[(iV^{\rm diss}(\boldsymbol r,t)+V^{\rm box}(\boldsymbol r,t)) \psi_j(\boldsymbol r,t) ]\ \psi_j^*(\boldsymbol r,t){\big]},
\label{eq:bud_d}\end{eqnarray}
 where $\mathbb{F}[\cdots]$ is the Fourier operator.
Each component's cascade flux in a two-component BEC includes self-interaction $\boldsymbol\Pi^{(jj)}_n$ and inter-component interaction $\boldsymbol\Pi^{(jl)}_n$ ($j \neq l$) (Figure~\ref{f3}).

\begin{figure}
\centering
\includegraphics[width=16cm]{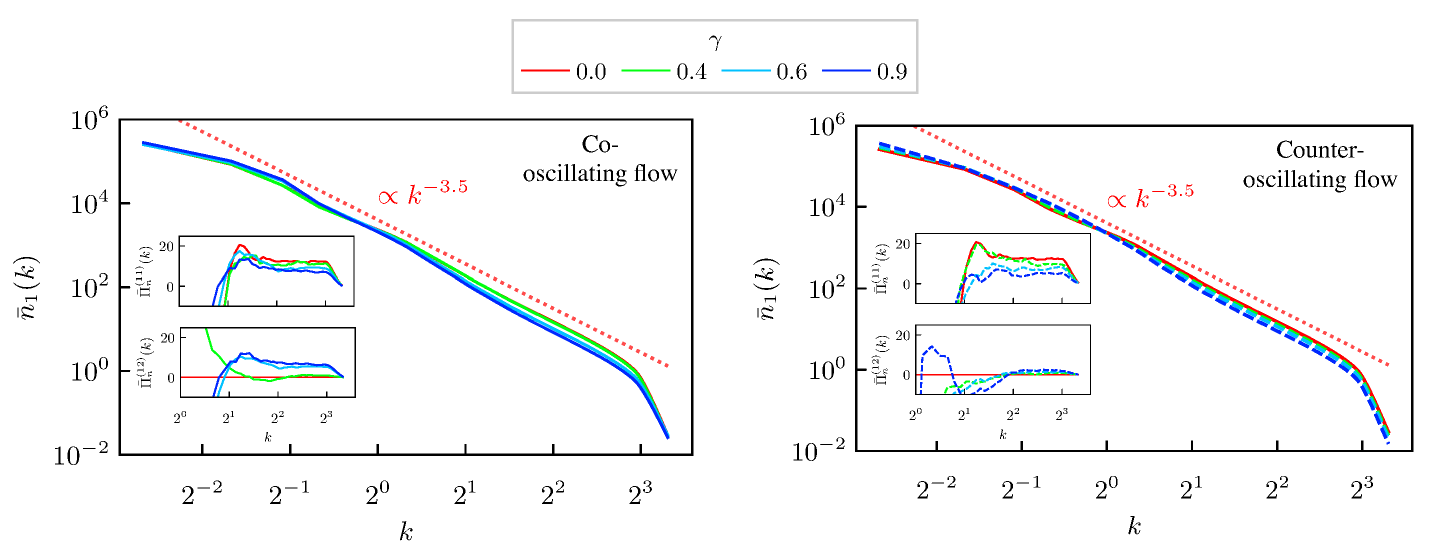}
\caption{Momentum distribution of component 1 at \( \gamma = 0, 0.4, 0.6, 0.9 \), with the solid line representing the co-oscillating flow and the dashed line representing the counter-oscillating flow in the quasi-steady state. The power law $\propto k^{-3.5}$ is indicated as references. The inset illustrates the wave number dependence of the particle fluxes for component 1, ${\Pi}^{(11)}_n(k)$ and ${\Pi}^{(12)}_n(k)$, in steady state.
}
\label{f4}
\end{figure}

The results for the co-oscillating flow cases ($\gamma = 0.0$, $0.4$, $0.6$, $0.9$) are shown on the left side of Figure~\ref{f4}, while those for the counter-oscillating flow cases are presented on the right. 
The momentum distribution for component 1 in the steady state was calculated as follows:
\begin{eqnarray}
\bar{n}_1(k)=\frac{1}{15T\times4\pi k^2}\int^{500+15T}_{500} \int_{|\boldsymbol k'|=k}n_1(\boldsymbol k', t) {\rm d}\boldsymbol k'{\rm d}t 
\end{eqnarray}
where $T = 2\pi / \omega_{\rm res}$.
Due to the symmetry between the two components, only the results for component 1 are displayed. 
Both co-oscillating and counter-oscillating flows exhibit the Kolmogorov-Zakharov spectrum \cite{Nir2016,Nir2019}, which follows a power law $\propto k^{-3.5}$, similar to single-component systems. This indicates that inter-component interactions do not disrupt the self-similar cascade process.

The inset of Figure.~\ref{f4} illustrates the wave number dependence of the particle fluxes for component 1. 
The particle flux in the steady state was calculated as follows:
\begin{eqnarray}
\bar{\Pi}^{(jl)}_n(k) = \frac{1}{15T} \int^{500+15T}_{500} \int^{\infty}_k \int_{|\boldsymbol{k''}|=k'} 2i {\rm Im} \big[ \mathbb{F}[g_{jl} |\psi_l(\boldsymbol{r},t)|^2 \psi_j(\boldsymbol{r},t)] \psi_j^*(\boldsymbol{r},t) \big] {\rm d} \boldsymbol k'' {\rm d}k' {\rm d}t. \label{eq:bud_nl}
\end{eqnarray}
In the high-wavenumber region, direct cascade fluxes independent of $k$ were observed, similar to the single-component system\cite{Nir2019,Nir2024}. For coupled turbulence, characterized by co-oscillating flows at $\gamma = 0.6$ or $0.9$, a significant increase in ${\Pi}^{(12)}_n$ indicates enhanced inter-component interactions. Conversely, in decoupled turbulence, observed in co-oscillating flows at $\gamma = 0.4$ or in all counter-oscillating flow cases, ${\Pi}^{(12)}_n$ was nearly zero, indicating a decoupling between the two components.

\end{document}